September 22, 2011

# Otto Stern (1888-1969):
# The founding father of experimental atomic physics

**J. Peter Toennies,**[1] **Horst Schmidt-Böcking,**[2] **Bretislav Friedrich,**[3] **Julian C.A. Lower**[2]

[1]Max-Planck-Institut für Dynamik und Selbstorganisation
Bunsenstrasse 10, 37073 Göttingen

[2]Institut für Kernphysik, Goethe Universität Frankfurt
Max-von-Laue-Strasse 1, 60438 Frankfurt

[3]Fritz-Haber-Institut der Max-Planck-Gesellschaft
Faradayweg 4-6, 14195 Berlin

We review the work and life of Otto Stern who developed the molecular beam technique and with its aid laid the foundations of experimental atomic physics. Among the key results of his research are: the experimental determination of the Maxwell-Boltzmann distribution of molecular velocities (1920), experimental demonstration of space quantization of angular momentum (1922), diffraction of matter waves comprised of atoms and molecules by crystals (1931) and the determination of the magnetic dipole moments of the proton and deuteron (1933).

**Introduction**

Short lists of the pioneers of quantum mechanics featured in textbooks and historical accounts alike typically include the names of Max Planck, Albert Einstein, Arnold Sommerfeld, Niels Bohr, Werner Heisenberg, Erwin Schrödinger, Paul Dirac, Max Born, and Wolfgang Pauli on the theory side, and of Konrad Röntgen, Ernest Rutherford, Max von Laue, Arthur Compton, and James Franck on the experimental side. However, the records in the Archive of the Nobel Foundation as well as scientific correspondence, oral-history accounts and scientometric evidence suggest that at least one more name should be added to the list: that of the "experimenting theorist" Otto Stern. With 81 nominations, Otto Stern was the most nominated candidate for the Physics Nobel Prize during the period from 1901 until 1950, with 7 nominations more than Max Planck and 15 more than Albert Einstein.[1]

In 1919, Stern conceived an experimental approach to measuring internal quantum properties of single isolated atoms. In 1922, jointly with Walther Gerlach, he implemented this approach in the laboratory and proved that *Richtungsquantelung* (space quantization), predicted on theoretical grounds by Arnold Sommerfeld[2] and Peter Debye[3] was not just a figment of the mathematician's imagination but that it really existed. The Stern-Gerlach experiment turned out to be one of the milestones on the winding road to modern quantum physics, one which offered other-than-spectroscopic evidence that quantum objects (atoms) exhibit behavior incompatible with classical physics.

At the core of the Stern-Gerlach experiment, carried out at the University of Frankfurt, was the so-called *Molekularstrahlmethode* (molecular beam method), which Stern and his coworkers had further advanced and made use of between 1923 and 1933 at the University of Hamburg. During the Hamburg period, Stern's group's experiments provided evidence for other key manifestations of the quantum nature of matter, such as diffraction of He-atom matter waves by a crystal surface or the anomalous magnetic moments of the proton and deuteron. In 1943, the Nobel Prize for Physics was awarded to Stern "for his contribution to the development of the molecular ray method and his discovery of the magnetic moment of the proton."

In subsequent decades, Stern's molecular beam method had been widely adopted by the physics and physical chemistry communities world wide, and about 20 Nobel Prizes were awarded for work based on the method, including that on the MASER, NMR, and the atomic clock.

Otto Stern was born on February 17, 1888 as the eldest child of the well-to-do Jewish miller and grain dealer Oskar Stern and Eugenie, nèe Rosenthal, in *Sohrau* (Zory) in Upper Silesia. The family moved in 1892 to *Breslau* (Wroclav) where Stern went to the humanistic *Johannes Gymnasium*. After the *Abitur* (high-school graduation) in 1906, Stern took 12 semesters of physical chemistry at the Breslau University. His PhD adviser was the *Privatdozent* Otto Sackur, who had derived, simultaneously with but independently of Hugo Tetrode, a quantum statistical expression for the entropy of a monoatomic gas. It was due to Sackur's influence that Stern developed an abiding interest in entropy, which he maintained throughout his life. Stern received his PhD in April 1912 with a thesis on a topic of his own choice, namely the osmotic pressure of carbon monoxide in highly-concentrated solutions.

**Assistant to Einstein 1912–1914**

After earning his PhD, Otto Stern joined in May 1912 Albert Einstein as his first postdoctoral student and co-worker. Sackur, who was friendly with Fritz Haber, asked the renowned physical chemist to weigh in and introduce Stern to Einstein, his personal friend. Stern himself considered joining Einstein an impudence, but went nevertheless. Einstein was then at Prague's German University, his first station as full professor. As Stern later recounted in his Zurich interview with Res Jost:[4]

> I expected to meet a very learned scholar with a large beard, but found nobody of that kind. Instead, sitting behind a desk was a guy without a tie who looked like an Italian roadmender. This was Einstein. He was terribly nice. In the afternoon he was wearing a suit and was shaven. I had hardly recognized him.

When Einstein, in October 1912, had accepted a call to the University of Zurich, he invited Stern to come along and appointed him his scientific assistant. During the Zürich period, Stern and Einstein published a joint paper *Einige Argumente für die Annahme einer molekularen Agitation beim absoluten Nullpunkt* (Some arguments in favor of the conjecture of a zero-point molecular motion)[5] which examined aspects of the problem of zero-point energy. Characteristic for Einstein and Stern is a footnote added to the paper, which reflects their unconventional way of thinking and open-mindedness (*Querdenken*):

> It hardly needs to be emphasized that our way of handling this problem is only justified by our lack of knowledge of the correct laws governing the resonators.

On June 26, 1913 Otto Stern submitted his application for *Habilitation* and *Venia Legendi* in the field of physical chemistry and later that year became a *Privatdozent* at Zurich.[6] His 8-page *Habilitationschrift* was entitled *Zur kinetischen Theorie des Dampfdruckes einatomiger fester Stoffe und über die Entropiekonstante einatomiger Gase* (Kinetic theory of the vapor pressure of monoatomic solids and of the entropy constants of monoatomic gases).

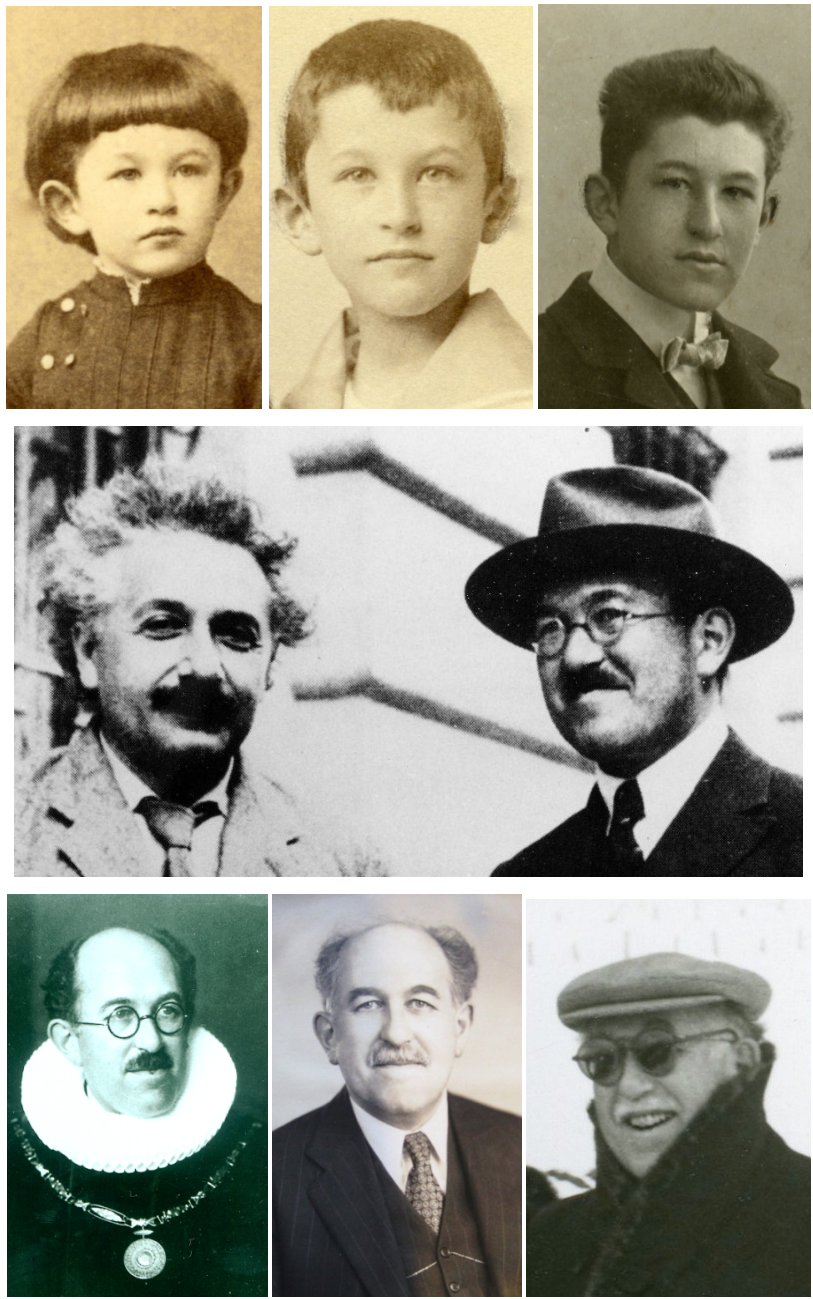

At Zurich, Stern met Max von Laue, who held a professorship at the Zurich University since 1912, the year of his momentous discovery of X-ray diffraction from crystals. Stern and von Laue shared, apart from what would become a life-long friendship, profound misgivings about the model of the atom proposed by Niels Bohr in 1913. Bohr's model combined the planetary model of the atom developed by Ernest Rutherford in 1911 with postulates about the angular momenta of the orbiting electrons and the light quanta emitted or absorbed when the electrons changed their orbits. In order to give an expression to their horror over the departure from classical physics Bohr's model entailed, Stern and von Laue vowed that "if this nonsense of Bohr should prove to be right in the end, we will quit physics." The vow was later dubbed by another scientist-in-residence at Zurich at the time, Wolfgang Pauli, the *Ütlischwur* – in a double-reference to a hill on Zurich's outskirts and the *Rütlischwur* from Friedrich Schiller's *Wilhelm Tell*.[8] Ironically, it was the Stern-Gerlach experiment which would prove that Bohr – or rather quantum mechanics – was right. In his Zurich interview, Stern added a twist to the story:[4]

> Einstein mentioned to me that he had thought about something like Bohr's atom himself. Well, Einstein was not as silly as we were.

Although Stern earned his *Habilitation* in theoretical physical chemistry, he was not a theorist by training. However, at both Prague and Zurich, he faithfully attended Einstein's lectures, which amounted to an apprenticeship in theoretical physics, whose intensity was enhanced by the absence of any other interlocutors than Stern during Einstein's stint in Prague. As Stern pointed out in his Zurich interview:

> Einstein never prepared his lectures. Einstein just improvised, but in a physically interesting and sophisticated way. […] I learned the "*Querdenken*" from him. […] I also learned from Einstein to talk nonsense every now and then. Einstein registered with pleasure when he had made a mistake. He would admit his mistake and remark: it's not my fault that *der liebe Gott* (the dear Lord) didn't make things the way I had imagined.

Immanuel Estermann, a close co-worker and friend of Stern's characterized the relation between Stern and Einstein as follows:[9]

> From his collaboration with Einstein, the real benefit was to learn how to distinguish which problems of contemporary physics were important and which were not so important; which questions to ask and which experiments to undertake in order to answer the questions. Thus from a brief scientific collaboration evolved a close, life-long friendship, which would be the basis for Stern's great achievements.

**Stern's beginnings in Frankfurt and the intervening Great War (1914–1918)**

In 1914, Max von Laue accepted a professorship in theoretical physics at the newly established *Königliche Stiftungsuniversität Frankfurt* (The Royal University of Frankfurt) and appointed his friend Otto Stern to the post of his assistant; this post was augmented when the Frankfurt University recognized Stern's Zurich habilitation and appointed him, as of November 10, a *Privatdozent* in theoretical physics (although, formally, Stern remained tied to Zurich until the end of 1915). After the outbreak of the First World War, Stern reported to the German Army as a volunteer and was sent to Berlin to train as a meteorologist, whereafter he served as a non-commissioned officer at army headquarters. Among his tasks was to assist a captain in building up a physics laboratory in Belgium. However, there was no physics equipment available, except for some air pumps. In order to be able to fulfill his mission nevertheless, the captain came up with the idea of dismantling the near-by Solvay Institute and confiscate its physics apparatus for the military laboratory. Otto Stern set out to prevent, at any cost, the scavenging of the Solvay Institute. He secretly contacted Walther Nernst in Berlin, who used his influence to preserve the Solvay Institute. As Stern recalled in his Zurich interview, the captain would come to Stern two weeks later to complain that "Berlin" had stopped his efforts.

From the end of 1915 on, Stern served as a meteorologist at a field weather station in Lomsha in Russian Poland. Since this job provided him with plenty of free time, he kept busy thinking about topics in thermodynamics, "in order to keep his sanity." While in Lomsha, Stern wrote two extensive papers on entropy. Throughout his life, Stern was no great letter-writer, often admitting that this was not his cup of tea. However, in Lomsha Stern seems to have deviated from his patterns. The collection of Otto Stern's family (Family Templeton-Killen) and the Bancroft Archive at Berkeley[10] hold several letters from the Lomsha period that Einstein and Stern had exchanged, which contain a discussion on the topics covered by Stern's Lomsha papers. However, as the correspondents were unable to come to the same conclusion, the papers were authored by Stern alone; Einstein's contribution to the published material was likely insignificant, as also attested by the absence of an acknowledgment of Einstein's help in the papers.

Alan Templeton, a grandnephew of Otto Stern, told one of the authors (HS-B) that a weather surveillance aircraft based in Lomsha had been shot down by the Russians with Otto Stern onboard. However, Stern survived the accident unscathed. This episode was the apparent reason for Stern's reluctance later in life to board a plane and for his predilection toward travel between America and Europe by ocean liners.

During World War I, many scientists, not just in Germany, had been engaged in military research. One of the centers of such research in Germany was Walther Nernst's laboratory at the Berlin University. Otto Stern joined Nernst's laboratory in November 1918, to work with the experimentalists James Franck and Max Volmer there. The three-month collaboration between Stern and Volmer resulted in three experimental papers on the kinetics of intermolecular deactivation processes, such as the quenching of fluorescence, governed by what is known today as the Stern-Volmer relationship.[11] More importantly, Stern's experience in Nernst's laboratory converted him from a theorist to an experimentalist.

**Back to Frankfurt (1919 – 1921)**

In order to help the veterans of WWI to catch up with their studies, the University of Frankfurt set up a trimester system, and Stern was called upon to give an introductory course on thermodynamics in the extra trimester running from February 3 until April 16, 1919.[12] By that time, Max von Laue had left Frankfurt and assumed an *Ordinarius* professorship at his alma mater, the Berlin University, side by side with his mentor Max Planck. Laue's move to Berlin was a part of a swap that brought Max Born from Berlin, where he held an *Extraordinarius* professorship, to Frankfurt, where he was appointed an *Ordinarius* for theoretical physics, thereby "inheriting" Otto Stern as his assistant. Born's Institute at Frankfurt consisted of another *Privatdozent*, Alfred Landé, and another assistant, Elizabeth Bormann. In addition, Born's Institute for Theoretical Physics also comprised a machine shop, run by a distinguished fine-mechanic, Adolf Schmidt, who proved instrumental for the later success of Born's small Institute. Here is how Born described Stern's beginnings at his Institute:[13]

> I was fortunate enough to have found in Otto Stern a Privatdozent of the highest quality, a good-natured, cheerful man, who had soon become a good friend of ours. The work in my department was guided by an idea of Stern's. He wanted to measure the properties of single atoms and molecules in gases by making use of molecular beams, which were first employed by Louis Dunoyer in 1911. Stern's first apparatus was designed to produce direct evidence for the velocity distribution law of Maxwell and to measure the mean velocity. I was so fascinated by the idea that I put all the means of my laboratory, workshop and mechanic at his disposal.

Max Born's institute in Frankfurt was not a big operation. Born:[13]

> I had only two rooms in Frankfurt. And in one room there were some students ... Stern's apparatus was made up in my little room, so I saw it from the beginning and watched. And I was quite envious of how he managed: he did not touch it at all, for he is also, just like me, not very good with his hands. But we had a very good mechanic [Mr. Adolf Schmidt] and he did it for him. He [Stern] told him what to do and it came out.

In his first benchmark experiment at Frankfurt, Stern set out to verify the Maxwell-Boltzmann distribution of the velocities of gaseous molecules. Stern had made use of a beam of atoms produced by heating silver to a given temperature. In this way, he had not only corroborated a theoretical result dating back to the 1860s by a cogent experiment, but also secured a future for the molecular beam method, first employed by Louis Dunoyer in 1911.[14] Dunoyer showed that sodium atoms travel in vacuum (at a pressure of about $10^{-3}$ millibar) along straight lines like light and produce a well-defined shadow image of an obstacle placed in their way, and thereby confirmed one of the key assumptions of the kinetic theory of gases.

A molecular beam consists of myriads of single atoms/molecules separated from one another by a distance large enough to preclude interactions among them. Therefore, a molecular beam in effect offers the possibility to experiment with single, isolated atoms or molecules. However, in order for the molecular beam experiments to be quantitative, the molecular beams have to be well characterized. The velocity distribution of the beam molecules is one such key characteristic, and so Stern's first Frankfurt experiment prepared the soil for much of what would come later in his own laboratory as well as in the laboratories of others who would implement the molecular beam technique.

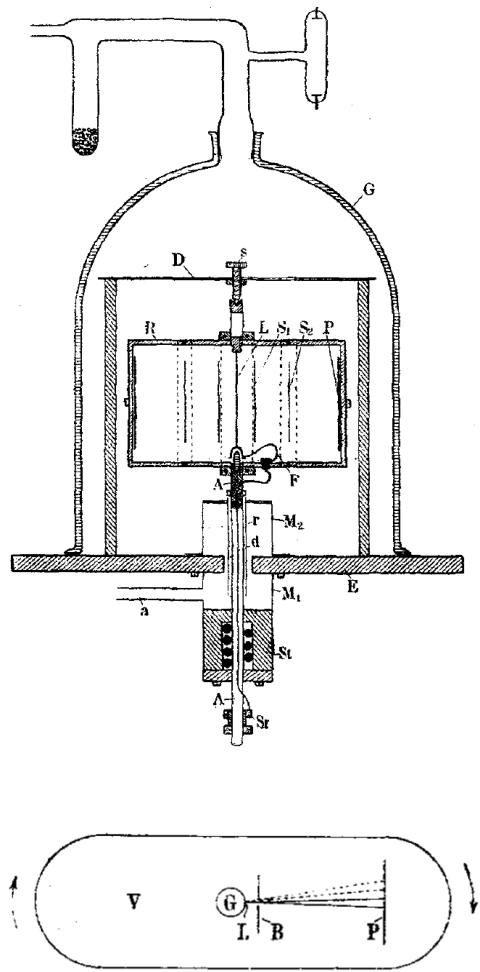

Fig. 2: Stern's apparatus for measuring the Maxwell-Boltzmann distribution of molecular velocities, in side-view (top) and top-view (bottom).[16]

Stern's simple, ingenious apparatus to determine the Maxwell-Boltzmann velocity distribution was designed as follows, see Fig. 2: An electrically-heated thin platinum wire (L) dipped into silver soldering paste served as a source of silver atoms, radially emitted into the gas phase by evaporation. The effusive beam was narrowly collimated by a pair of horizontal and vertical slits ($S_1$ and $S_2$) and, like in Dunoyer's experiment, captured on a cold plate (P), where the atoms condensed and turned visible. The position and shape of the condensate spot could be accurately measured with a microscope. The slits and the cold plate were mounted on a platform which could be rotated around the beam source at a rate of up to 2,400 rotations per minute. When rotated, the Coriolis force acting on the silver atoms in the rotating frame of the platform resulted in a shift of their position on the cold plate with respect to the position at which they impinged when the platform was at rest. From the measured shift, the geometry of the apparatus, and the speed of rotation it was then possible to determine the mean radial velocity of the atoms. Although Stern originally intended to determine the entire velocity (speed) distribution, he submitted in April 1920 a paper to *Physikalische Zeitschrift* entitled *Eine direkte Messung der thermischen Molekulargeschwindigkeit* (A measurement of the thermal molecular speed)[15] in which he only reported the atoms' mean velocity. This came out as 650 m/s for an assumed source temperature of 961°C.[15] However, the Maxwell-Boltzmann formula predicted 534 m/s, about 15% less, which Stern first blamed on the uncertainty in determining the source temperature. The real cause of the discrepancy was pointed out to Stern by his mentor Einstein: Stern omitted a key transformation required when the molecules pass through a slit – in which case their transmission differs for different molecular velocities. By applying the transformation to the Maxwell-Boltzmann distribution, Stern obtained a mean velocity greater by a factor of 15% than before, and was able to enjoy its gratifying agreement with his old data as well with new measurements he undertook at different source temperatures and rotation rates.[16] Stern's concluding remark aptly captures the significance of the work reported:

> The experimental set up employed herein makes it possible to prepare, for the first time, molecules with a uniform velocity.

This proved to be a key moment for physics: from then on it would be possible to prepare isolated molecules in a well-defined momentum state and thus to accurately measure any momentum changes imparted to the molecules by external fields or other molecules. Stern's first Frankfurt experiment was a milestone on the path to quantum physics.

With his training in theory by Einstein, Stern was able to conceive the most imaginative ideas for experiments and experimental apparatus. However, as he admitted in his Zurich interview, he lacked the skills and the dexterity needed to implement them in the laboratory, at least at the beginning of his Frankfurt time. Therefore, he sought the help of an experienced experimentalist, whom he found in his Frankfurt colleague Walther Gerlach. The Stern-Gerlach experiment they carried out together in 1921-1922

> ranks among the dozen or so canonical experiments that ushered in the heroic age of quantum physics. Perhaps no other experiment is so often cited for elegant conceptual simplicity. From it emerged both new intellectual vistas and a host of useful applications of quantum science.[17]

In their first and last joint venture, Stern and Gerlach undertook to find out whether the so called "space quantization" was real. The idea of space quantization was developed in 1916 nearly simultaneously but independently by Arnold Sommerfeld[2] and Peter Debye[3] in an attempt to amend Niels Bohr's 1913 model of the atom to account for the normal Zeeman effect, i.e., the splitting of spectral lines of (hydrogenic) atoms by a magnetic field. Whereas the anomalous Zeeman effect (which arises for atoms in other than singlet states) would baffle atomic physicists for one more decade, until the discovery of electron spin in 1925, space quantization as an archetypal manifestation of the quantum world would remain striking up to the present …

Classically, the atomic magnetic moments could be oriented at an arbitrary angle with respect to an external magnetic field. In contrast, Sommerfeld's and Debye's idea amounted to postulating that the magnetic moment could only take certain discrete orientations with respect to the field – that its direction is "spatially quantized" and not "classically continuous." To add to the strangeness, the discrete orientations of the magnetic moments were to change if their "observer" picked another direction of the external magnetic field. Even Debye himself did not believe in the reality of space quantization and confided his misgivings to Gerlach:[18]

> You surely don't believe that [space quantization] is something that really exists; it is only a computational recipe, a timetable of the electrons.

Max Born let his voice be heard (a little later) as well:[13]

> I always thought that space quantization was only a symbolic expression for something you don't understand.

And Otto Stern, according to his Zurich interview, did not believe in the existence of space quantization either. He wanted to prove that the whole concept was flawed.

On August 26th 1921, Stern submitted a paper to *Zeitschrift für Physik* entitled *Ein Weg zur experimentellen Prüfung der Richtungsquantelung im Magnetfeld* (A way towards the experimental examination of spatial quantization in a magnetic field) in which he described how to test whether space quantization was for real. As he put it,[19]

> Whether ... the quantum theoretical or classical interpretation is correct can be decided by a basically very simple experiment. One only needs to investigate the deflection which a beam of atoms experiences in an appropriate inhomogeneous magnetic field.

Stern's superior at the time, Max Born, recalled later in his interview with Paul Ewald:[20]

I tried to persuade Stern that there was no sense, but then he told me that it was worth a try.

Stern himself expressed in the paper his own misgivings about space quantization:[19]

> A further difficulty for the quantum interpretation, as has already been noted from various quarters, is that one just cannot imagine how the atoms of the gas, whose [magnetic moments] without magnetic field have all possible directions, are able, when brought into a magnetic field, to align themselves in the pre-ordained directions. Really, something completely different is to be expected from the classical theory. The results of the magnetic field, according to Larmor, is that all the atoms perform an additional uniform rotation with the direction of the magnetic field strength as axis, so that the angle which the direction of the [magnetic moment] makes with [the magnetic field] continues to have all possible values for the different atoms.

There is a footnote added to Stern's paper which provides an explanation as to why Stern published about an experiment much of which was yet to be done: Hartmut Kallmann and Fritz Reiche, based at Fritz Haber's Kaiser Wilhelm Institute in Berlin, had submitted a paper on closely related work. Although Kallmann's and Reiche's goal was different, namely to test whether the electric dipole moment of polar molecules was an individual or a bulk property, there had been a considerable methodological overlap between their and Stern's and Gerlach's work, and Stern sought to make this known:[19]

> Mr. Gerlach and I have been occupied for some time with the realization of [the Stern-Gerlach] experiment. The reason for the present publication is the forthcoming paper by Messrs. Kallmann and Reiche concerning the deflection of electrical dipolar molecules in an inhomogeneous electric field.

In early 1921, Stern and Gerlach started working in earnest on the design and execution of their experiment to test the concept of space quantization. Technically, this was a difficult experiment to carry out. The molecular beam part of the apparatus had to be rather small – not much bigger than a fountain pen – in order to fit into a glass vacuum chamber, and likewise restricted by the size of the electro-magnet. This core of the apparatus was subject to a large temperature gradient, as the beam source – a silver oven – had to be heated to about $1,300^{o}C$ and the Gaede mercury diffusion pumps, used to generate a vacuum, as well as the condenser plate had to be cooled to the temperature of liquid air . At the same time, the set up (oven, slits, magnet, condenser plate) had to be very accurately aligned, as the deflection of the silver atom beam by the inhomogeneous magnetic field was expected to be only on the order of 0.1 mm, and an inaccuracy in the alignment on the order of 10 μm could not be tolerated. Because of the small intensity of a molecular beam, the experimental runs took several hours, during which the delicate apparatus had to remain aligned and otherwise well behaved.[21]

Apart from a shortage of funding,[22] the experiment was hindered by other external circumstances as well. In May 1920, Max Born received a call to the University of Göttingen. However, since he was quite happy in Frankfurt, he wanted to stay. In a letter of June 7, 1920 addressed to the city's Mayor Georg Voigt,[12] Born made five requests whose fulfillment would have kept him put. All of them were granted, except for the one which mattered to Born the most: to appoint Otto Stern as Professor at Frankfurt. On June 10, 1920, Born wrote to Voigt:

> Unfortunately, it seems impossible to fulfill my main wish, namely to attach my co-worker, Prof. Stern, to Frankfurt through a professorship. Exactly this point, namely the recruitment of outstanding faculty, is handled much more favorably by the Ministry in Göttingen.

As a result, in 1921 Born moved to Göttingen. In the Fall of 1921, Stern too received an offer, to become *Extraordinarius* (associate professor) for theoretical physics at the University of Rostock. In the winter semester 1921/22, he was already lecturing there on the subject. Hence, since autumn 1921, the Stern-Gerlach experiment was run by Walther Gerlach alone.

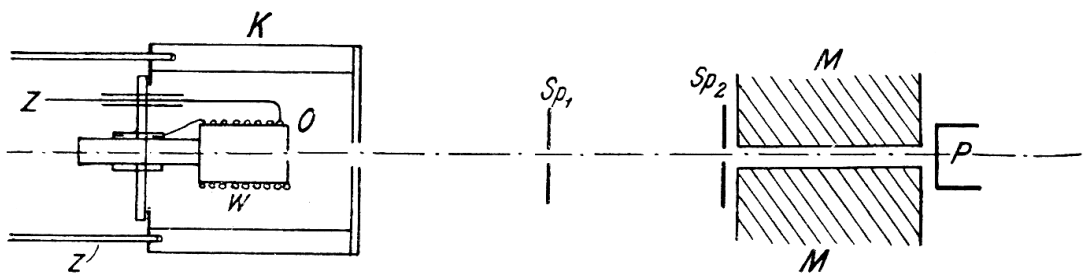

Figure 3: Schematic of the set up used in the Stern-Gerlach-experiment: Silver oven (*O*), collimator slits (*Sp₁* and *Sp₂*), inhomogeneous magnetic field (*M*), and P condenser plate (*P*).

Fig. 3 shows a schematic of the set up used for the Stern-Gerlach experiment.[23] An effusive beam of silver atoms, produced by heating silver metal in an oven, was collimated by a pair of slits, passed along the sharp pole piece of the magnet, and detected on a condenser plate attached behind the magnet. The total length of the apparatus (including the oven) was about 12 cm. Fig. 4 shows a photo of the apparatus, with part of the Hartmann & Braun magnet (a small Dubois magnet) removed, enabling to see the key components. The glass bell-shaped vacuum chamber on the left housed the oven. The chamber was connected to a vacuum pump and its double walls cooled by refrigerated air. The slits as well as the sharp pole piece were located behind the white quadratic structure in a vacuum tube (center). The condenser plate was housed in the cooling cylinder on the right.

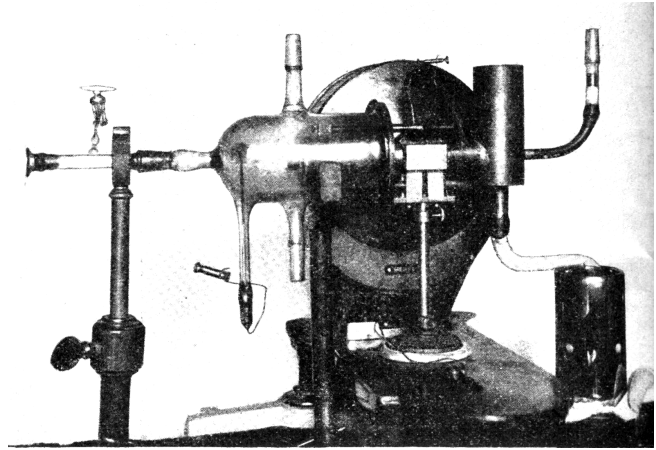

Figure 4: Stern-Gerlach apparatus.

Because Stern was required to teach in Rostock, he could visit Frankfurt only during breaks such as Christmas (1921) and Easter holidays (1922). The apparatus had been steadily improved by Gerlach and during a run during the night of November 5, 1921, Gerlach scored his first great success. A silver beam of a 0.05 mm diameter collimated by a pair of circular apertures 3 cm apart was dispersed by an inhomogeneous magnetic field acting upon the beam over a length of 3.5 cm in a vacuum of about $10^{-5}$ millibar, leaving behind a broadened spot on the condenser plate.[24] The width and shape of the spot allowed to infer that the internal magnetic moment of the silver atoms had a magnitude between 1 and 2 Bohr magnetons ($\mu_B$). On November 18, 1921, Gerlach and Stern submitted this result to *Zeitschrift für Physik*.[25] Because of the limited angular resolution, the outcome of the experiment would remain inconclusive as to the issue of space quantization in general, and the details of its manifestation in particular: Debye and Sommerfeld had predicted a triplet structure, in analogy to the normal Zeeman splitting, for the detected silver beam, with one component deflected downward, one upward, and one undeflected. Bohr, in contrast, predicted a doublet splitting, corresponding to an electron orbiting around the atom nucleus clock-wise or counter-clockwise with respect to the direction of the magnetic field and resulting in two components, one deflected upward and one downwards, but no undeflected beam.

Wilhelm Schütz, who at the time was a PhD student under Gerlach, allowed to watch Gerlach at work, has provided a testimonial of what he saw:[26]

> Anyone who has not been through it cannot at all imagine how great were the difficulties with an oven to heat the silver up to a temperature of 1300°C within an apparatus which could not be fully heated [the seals would melt] and where a vacuum of $10^{-5}$ torr had to be produced and maintained for several hours. The apparatus was cooled with dry ice and acetone or with liquid air. The pumping speed of the Gaede mercury pumps or the Volmer mercury diffusion pumps was ridiculously small in comparison with the performance of modern pumps. And then, their fragility; the pumps were made of glass and quite often they broke, either from the thrust of boiling mercury - despite an addition of lead - or from the dripping of condensed water vapor. In that case the several-day effort of pumping, required during the warming up and heating of the oven, was lost. Also, one could by no means be certain that the oven would not burn through during the four- to eight-hour exposure time. Then both the pumping and the heating of the oven had to be started from scratch. It was a Sisyphus-like labor and the main load and responsibility was carried on the broad shoulders of Professor Gerlach. In particular, W. Gerlach took over the night-guards. He would get in about 9 p.m. equipped with a pile of reprints and books. During the night he then read the proofs and reviews, wrote papers, prepared lectures, drank plenty of cocoa or tea and smoked a lot. When I arrived the next day at the Institute, heard the intimately familiar noise of the running pumps, and found Gerlach still in the lab, it was a good sign: nothing broke during the night. … I arrived one morning in February 1922 at the institute; it was a gorgeous morning; cold air and snow! W. Gerlach was in the middle of developing the silver deposit left by an atom beam which ran for 8 hours through the inhomogeneous magnetic field. Full of expectation, we watched the development process and have experienced the success of many months of hard work: the first splitting of a silver beam by an inhomogeneous magnetic field. After Meister Schmidt and, if I remember correctly, also E. Madelung, saw the splitting, the image was recorded micro-graphically in the mineralogical Institute. Then I got the job to send a Telegram to Herrn Professor Stern in Rostock, which read "Bohr is right after all!" ("*Bohr hat doch recht!*").

Wilhelm Schütz's apt description of the weather conditions "the morning after" made it possible for us to unambiguously date the fateful night when space quantization was demonstrated: a comparison with the records of the *Wetteramt* (Weather Service) attests that it was the night from 7 to 8 of February, 1922.[27] The above-mentioned telegram sent to Stern has probably been lost; at least it is not present in the Stern collection at the Bancroft Archive.

However, a postcard dispatched to Niels Bohr has been preserved, see Fig. 5. It was sent on February 8, as can be discerned by inspecting the right bottom corner of the rear side of the card with the silver beam deflection pattern.

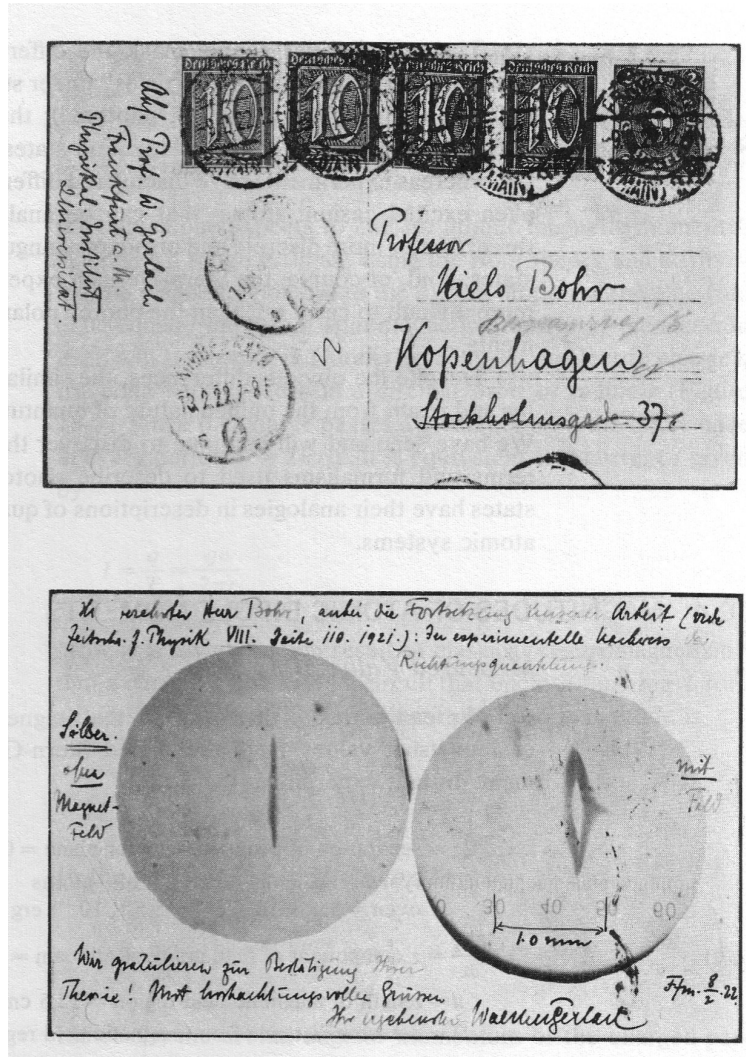

Figure 5: Gerlach's postcard to Niels Bohr, dated February 8, 1922.
Courtesy of the Emilio Segré Visual Archives.

Gerlach's postcard message to Bohr reads:

Hochverehrter Herr Bohr, attached a sequel to our work (see Zeitschrift f. Physik VII, page 110, 1921). The experimental proof of space quantization (silver without and with magnetic field). We congratulate you on the confirmation of your theory. With respectful greetings faithfully yours Walther Gerlach (Frankfurt, February 8, 1922).

On March 1, 1922, Walther Gerlach and Otto Stern submitted their landmark paper *Der experimentelle Nachweis der Richtungsquantelung im Magnetfeld* (The experimental proof of space quantization in magnetic fields) to *Zeitschrift für Physik*, with the paper's main message printed with emphasis:[28]

> Space quantization in a magnetic field has been proven as a fact.

The paper featured the same images of the split and unsplit silver beam as the postcard to Bohr did: without a magnetic field there is no deflection of the beam; with an inhomogeneous magnetic field applied there is a kiss-like spot affirming a splitting of the silver beam. In the experimental run that led to the final success, Gerlach used a 0.8 mm long and 0.03 mm wide platinum slit instead of a circular aperture. Since the slit's long side was oriented perpendicular to the direction of the magnetic field, only the fraction of the beam passing directly underneath the sharp edge of the magnetic pole piece (where the field strength was largest) showed a splitting. The splitting of the beam at the center of the slit clearly revealed a doublet structure. Thereby Sommerfeld's and Debye's concept of space quantization was corroborated.

The Stern-Gerlach experiment had thus unambiguously demonstrated that space quantization was not a mathematical artifact, but real physics. The finding that the atoms "knew" the direction of the magnetic field that the experimentalist has randomly chosen was deplored as particularly puzzling. Einstein together with Paul Ehrenfest tried in vain to find a possible classical explanation for the Stern-Gerlach effect and showed that a classical mechanism for orienting the magnetic dipoles, such as externally-induced radiative processes, was altogether lacking.[29]

We know today that aside from possible stray fields, it was solely the direction of the Stern-Gerlach magnetic field that provided the reference frame for space quantization and that there is no classical analog for the entanglement between the dipoles and the field direction. Gerlach and Stern had been the first to provide evidence for this entanglement.

Another puzzling issue was the splitting pattern observed in the Stern-Gerlach experiment. Since the data seemed to conform to Bohr's prediction, the prevalent perception was that Bohr was "right" and Sommerfeld and Debye "wrong." In hindsight, neither was "right," since the magnetic moment of the silver atom was not due to the electron's orbital angular momentum (moreover incorrectly presumed as having the value of h/2π)[30] but rather to the electron's "internal" angular momentum, i.e., spin, which would be discovered only three years later. However, in the early 1920s, there was one scientist who could have sounded an alarm and pointed out that the Stern-Gerlach doublet splitting was hinting at something other than the electron's orbital degree of freedom. But he kept quiet. This researcher was Alfred Landé, who had received his *Habilitation* under Born in 1919 on the theory of many-electron atoms and was sharing with Stern the premises of Born's Institute at Frankfurt.

Landé published in 1921 a semi-empirical formula, which described quantitatively both the normal and the anomalous Zeeman splitting of atomic levels.[31] His key idea was that both the normal and the anomalous Zeeman effects have the same origin. By playing with experimental values, he found his famous g-factor formula relating the magnetic dipole moment of an atom to its angular momentum, which is identical with the correct quantum mechanical result. Had Landé considered the Zeeman and the Stern-Gerlach splitting as mutually related, he could have discovered the spin degree of freedom already in 1922 at Frankfurt. Here is how Max Born described Landé's work in the Ewald interview:[20]

> As far as I remember the main indications of the crisis were the multiplets and the Zeeman effect, and these things. That we called the zoology of terms. Landé came to my department – I don't know the period exactly – and was my student in Gottingen. […] Then he came to Frankfurt again, and his head was completely occupied with the paper which I did't grasp at first. It was these whole number relations between the intensities of multiplet lines and Zeeman-effect lines. And he did it in a way which seemed to me horrible, namely, simply by guessing about numerical values. He wrote long lists of numerical values and said they must be contained in one formula – how can one construct it? And he tried the most impossible things. And at last it came out. At last came a formula which gave all the results he wanted. I couldn't check it – I can never do numerical calculation problems. So I didn't take much notice of him, and he also did not take much notice of our work, though we were sitting all the time in the same room. But two years later, or three, when we derived the square root of integers formula from quantum mechanics, we saw at once that it was very important. Some of these formulae were known before for multiplets from the Dutchman Ornstein. But for the multiplets, I think, and the expression of this "g" were first given by Landé.

According to Landé's formula, electrons with angular momentum h/2π would assume three orientations with respect to a magnetic field, in agreement with the prediction of Sommerfeld and Debye. However, Landé's formula is also compatible with a doublet splitting, provided the Zeeman effect is due to an angular momentum of ½ h/2π with a g-factor of 2. Classically, it was impossible to explain the source of such a half-integral angular momentum: since an electron on a Bohr orbit could never have it, it would have to be a property of the electron itself. However, only in 1925 would Landé's former student assistant Ralph de Laer Kronig,[32] working at the time at Columbia University towards his PhD, and a few months later, the Dutch physicists George Eugene Uhlenbeck and Samuel Abraham Goudsmit[33] have the audacity to talk in public about the electron's "internal" angular momentum. In the case of Kronig, this had the unfortunate consequence of him being dissuaded by one of his interlocutors, Wolfgang Pauli, from pursuing the idea of electron spin any further:

> Kronig would have found the spin, had not Pauli frightened him. (*Der Kronig hätt den Spin entdeckt, hätt Pauli ihn nicht abgeschreckt*).

Although a proponent of the "fourth quantum number" for an atomic shell electron already since 1924,[34] Pauli was himself "frightened away" by an apparent incompatibility of a spinning electron with special relativity theory. However, this incompatibility was not there, as shown in 1926 by Llwellyn Hilleth Thomas,[35] whose relativistic analysis put the heuristic concept of a spinning electron on a firm footing. The hypothesis that

> the electron itself, spinning like a tiny gyroscope, is probably the ultimate magnetic particle was expressed already in 1921 by Arthur Compton, based on evidence from X-ray diffraction by magnetic crystals and the curvature of the tracks of beta rays through air[36] and was taken into account by Uhlenbeck and Goudsmit when they introduced their electron spin hypothesis. Apparently, at that time no one discussed Comtpon's hypothesis in connection with either the Zeeman or the Stern-Gerlach effect.

What seems particularly puzzling today is that neither in 1922 nor in 1925 had the Stern-Gerlach experiment been discussed in terms of electron spin.[17] The first mention that it was in fact electron spin which was responsible for the magnetic deflections observed in the Stern-Gerlach experiment appeared as late as 1937 as an aside in the second edition of Ronald Fraser's book.[37]

In 1922 much of the physics community, including Stern, was astonished by the experimental proof that space quantization existed. Pauli wrote to Gerlach:[18]

> This should convert even the nonbeliever Stern.

Sommerfeld provided this comment on the outcome of the Stern-Gerlach experiment:[18]

> Through their clever experimental arrangement Stern and Gerlach not only demonstrated the space quantization of atoms in a magnetic field, but they also proved the quantum origin of electricity and its connection with atomic structure.

Einstein wrote:[18]

> The most interesting achievement in quantum physics at this point is the experiment of Stern and Gerlach. The alignment of the atoms without collisions or via radiation cannot be explained by existing theory; it should take the atoms more than 100 years to become aligned. I have done a little calculation about this with [Paul] Ehrenfest.

Stern himself expressed his feeling about the experimental result in his 1961 Zurich interview:[4]

> I was unable to understand anything about the outcome of the experiment, the two discrete beams. It was totally incomprehensible. It is obvious [today] that [in order to comprehend the experiment] one needs not only the new quantum theory but also a magnetic electron. These are the two things which were still missing at the time. I was fully confused and did not know what to do with such a result. Even today, I have objections against the beauty of quantum mechanics. But it is correct.

During Easter break of 1922 Otto Stern came to Frankfurt to work with Gerlach on improving the quantitative aspects of the Stern-Gerlach experiment. Special attention was paid to determining accurately the inhomogeneity and strength of the magnetic field employed to impart a deflection to the silver atoms. As they described in their paper *Das magnetische Moment des Silberatoms* (The magnetic moment of the silver atom)[38] this was measured by

> weighing the repulsive force of a very tiny probe made out of bismuth from point to point [of the field] and the measurement of the field strength from the variation of the resistance of a thin bismuth wire strung parallel to the sharp edge of the pole piece.

The alignment technique as well as the design of the magnetic pole pieces was apparently proposed to Stern and Gerlach by Born's successor at Frankfurt, Erwin Madelung. In his Zurich interview, Stern expressed regret that the acknowledgment of Madelung's help in the paper was not more emphatic.

By taking into account the geometry of the apparatus, the inhomogeneity of the magnetic field, and the mean velocity of the beam atoms, Stern and Gerlach had found that the value of the magnetic dipole moment of the silver atoms was equal, within 10%, to one Bohr magneton. This appeared to be in gratifying agreement with the available Bohr-Sommerfeld-Debye theory. However, this agreement was only fictitious, brought about by an "uncanny conspiracy of Nature,"[17] consistent with the submission of the paper on the 1[st] of April: the anomalous gyromagnetic ratio of the electron ($\approx 2.0023$) roughly canceled the electron spin of of ½. But there was no way to tell. Thereby ended Stern's stint at Frankfurt.

**Stern in Rostock (1921–1922)**

Otto Stern's appointment at Rostock was connected with little funding, a heavy teaching load and, after the retirement of his only colleague, the professor of experimental physics, also time consuming administrative duties. However, it was at Rostock where Stern was joined by Immanuel Estermann, who would work with Stern until Stern's retirement in 1946. Stern's Rostock period was brief: in the fall of 1922 he received – and accepted – an offer for a full professorship from the University of Hamburg.

Stern's position in Rostock was held by a string of first-class physicists: Wilhelm Lenz (1920-1921), Walter Schottky (1923-1927), Friedrich Hund (1927-1928) and Pascual Jordan (1929-1944).

**Otto Stern's Golden Years in Hamburg (1923 – 1933)**

On January 1, 1923 Otto Stern took up his new position as Professor (*Ordinarius*) of  Physical Chemistry and Director of the Institute for Physical Chemistry at the University of Hamburg, which was founded shortly before, in 1919. The following ten and a half years in Hamburg were Stern's most successful, golden years of his research career.  In Hamburg Stern established an outstanding research group which through many spectacular pioneering contributions soon achieved world-wide fame and became the leading international center for atomic, molecular and nuclear physics. In 1926 he published the first in a series of 30 remarkable papers which were all subtitled *Untersuchung zur Molekularstrahl-Methode,* UzM (Investigations by the molecular-beam method). In the first of two visionary introductory articles[39] Stern discussed for the first time all the special advantages of the molecular beam method and laid out a program for future research with 8 major scientific goals, all of which were far ahead of their time. Among these were such boldly ambitious projects as measuring nuclear magnetic moments, which he estimated to be only "about 1/2000" of a Bohr magneton, in a Stern-Gerlach type of experiment; determining Einstein's photon recoil; and confirming Louis de Broglie's 1924 prediction of wave-particle duality. In the introduction he proclaimed that

> The molecular beam method must be made so sensitive that in many instances it will become possible to measure effects and tackle new problems which presently are not accessible with known experimental methods.[39]

He was fully aware that the molecular beam method stood in direct competition with optical spectroscopy, but in contrast to spectroscopy, which can only observe differences in the energies of two states of a given molecule, the molecular beam method can measure physical properties of an isolated molecule in a specified quantum state. In the second far-sighted article, published with Friedrich Knauer,[40] Stern described in detail how to produce highly collimated intense molecular beams required to increase precision while, at the same time, greatly reducing the measuring times of molecular beam experiments. He proposed to replace the thin (0.4 mm dia.) silver-coated platinum wire evaporator, used as the beam source in his 1920 measurements of velocity distributions, by a heated source chamber, which he called an "oven." With this new source the beam exit slit could be narrowed without losing intensity since, according to the Knudsen condition, it would be possible to raise the source pressure and thereby compensate fully for the loss of intensity due to a reduced transmission. Thanks to the increased intensity the deposited beam would be detectable with a chemical developer after only 3 to 4 seconds. With these and other measures, Stern and Knauer predicted on the basis of extensive numerical calculations that the angular and momentum resolution could be improved to the extent that the resolution of the magnetic moment in magnetic deflection experiments could be increased to 1 part in 100,000 of a Bohr magneton, more than sufficient for measuring even nuclear magnetic moments. He also predicted that the increase in sensitivity and angular resolution should make it possible to detect the scattering from surfaces or even from gases and in this way to determine the corresponding van der Waals forces.

Up to 1929, Stern's laboratory in Hamburg consisted of four rooms in the basement of the Physics Institute, which provided reasonable conditions by the standard of those times. Then in 1929 Stern was offered a position as Professor of Physical Chemistry at the University of Frankfurt.[41,42] Since the city of Hamburg and his Hamburg colleagues were quite anxious to keep Stern, the ensuing negotiations led to a significant improvement in his working conditions. Despite the hard times in Germany resulting from the global financial crisis of 1929, which triggered the Great Depression, he was offered a brand new building, additional staff positions and more than ample funds for the workshop and technicians. This explains why in 1930, when Max von Laue offered him a highly prestigious professorship at the Kaiser-Wilhelm Institute for Physical Research in Berlin, Stern felt strongly committed to Hamburg and turned down Laue's offer.[41] Undoubtedly the excellent working conditions in Stern's laboratory significantly contributed to its scientific successes in the remaining four years in Hamburg. During this period, Stern's research staff consisted of four assistants, a large number of foreign research fellows and four to five PhD students. His closest assistant was Immanuel Estermann who had come shortly before Stern as a *Privatdozent* from Rostock and would later emigrate with him to Pittsburgh. In Hamburg the group was joined by Friedrich Knauer, Robert Schnurmann and later in 1930 by Otto Robert Frisch, the nephew of Lise Meitner. The short, three-year collaboration with Otto Frisch was extremely rewarding for both Frisch and Stern. In his book "What little I remember,"[43] Frisch describes his first impressions after arriving in Stern's laboratory as follows:

> My first recollection of the laboratory is standing in front of what looked like a forest of glass, a sort of glass blowers nightmare; tubes and bulbs and cylinders and mercury pumps blown from glass, with stopcocks by the dozen connected in a manner that made no more sense to me than the twigs in a hedge. And there I watched Stern and his chief assistant, Immanuel Estermann, turning stopcocks apparently at random, closing this one and then after a few seconds opening that one, and so on for what seemed like half an hour. I felt I would no more learn this than a totally unmusical person would ever learn to play the organ.

With great candor Frisch goes on to describe Stern's way of doing experiments:

> Stern was rather clumsy, and moreover one of his hands invariably held a cigar (except when it was in his mouth); so he was disinclined to handle any breakable equipment and always left that to his assistants. I still remember what he did when anything appeared to topple. He would never try to catch it; he lifted both hands in a gesture of surrender and waited. As he explained to me: You do less damage if you let the thing fall than if you try to catch it. Yet Stern was, in a higher sense, a superb experimenter. In using a new apparatus he left nothing to chance. Everything had been worked out beforehand and every detail of the performance was carefully checked. Stern would calculate, for instance, how much beam intensity he expected to get, even though that was a very lengthy and tedious calculation, which he always did himself. He could not predict the intensity very accurately; but if it felt short by more than 30% he felt something must be wrong, and fault to track it down. I have never seen anybody keeping such strict control of his instruments, and it surely paid off. As a rule the experiments we did were so difficult that nobody else in the world was attempting them. That created an oddly relaxed atmosphere.

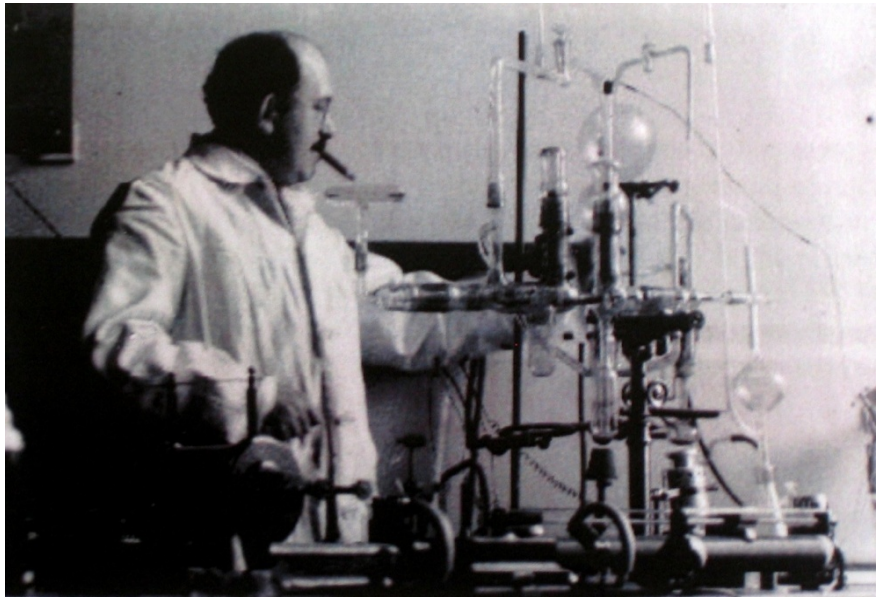

Figure  6: Otto Stern at work in his Hamburg laboratory.

The research fellows came mostly from Italy, England and the United States. Two of them, Isidor Isaac Rabi (U.S.A) and Emilio Segré (Italy), would later become especially famous and would be awarded Nobel Prizes for research which was inspired by their stays with Stern in Hamburg. Isidor Rabi had come to Hamburg to work with Wolfgang Pauli. In an interview in 1988 shortly before he passed away, Rabi told the science writer and author John Rigden how Otto Stern and his experiments affected his career:[44]

I first met Stern in the fall of 1927. I had been in Copenhagen at the Niels Bohr Institute of Theoretical Physics and Bohr [who at the time had too many visitors] made an arrangement for Yoshio Nishina and me to go and work with Wolfgang Pauli at the University of Hamburg. When I got there, I was pleased to find that Stern and his associates were engaged in very exciting molecular beam experiments. While my prime interest was with Pauli in theory, I spent time in Stern's laboratory talking with Ronald Fraser, a Scotsman, and John Taylor, an American. I came to understand the subtleties of the molecular beam experiments and recognized that the components of an atomic beam could be separated with a homogeneous [in place of the inhomogeneous field used in the Stern-Gerlach experiment] magnetic field. I explained the idea to Stern and he suggested do the experiment. I was told what an honor it was to be invited by Stern to do an experiment in his laboratory. I had no job and I had a wife to support. I was in no position to refuse the honor. My experiment was a success and when it came time to write up the results, I saw a demonstration of Stern's generosity, his fairness, and his pride. "First, publish a letter in Nature", said Stern. "If you publish it first in German, they'll think it's my thing, and it's yours.

In fact, this was the rule with Stern. Of the 30 articles in the UzM series 16 were published under the name of a single author without Stern's name. These included the PhD dissertations of Alfred Leu, Erwin Wrede, Berthold Lammert and Lester C. Lewis.[45] In 1944, Rabi was awarded a Nobel Prize in Physics "for his resonance method for recording the magnetic properties of atomic nuclei."

In an earlier interview with Thomas S. Kuhn in 1963, Rabi described the daily scientific life in Stern's Institute in the following way:[46]

I got to work and shared a lab with Taylor, who really taught me the technique. I saw very little of Stern himself, during that time. I did the experiment. All the time Walter Gordon was there, and later on Jordan came, and, of course, there was Lenz who was the professor; there was Pauli, and Bohr used to come, and Born. It was a place where people were in and out all the time. And of course there was Stern. The seminars were marvelous and the colloquium was very interesting, very high level, in the sense that there were different kinds of minds; Lenz, for instance, had a mind like a steel trap. He could make up things on the spot, although he never accomplished very much. Then there was Stern with his marvelous physical intuition and point of view, and Pauli with his tremendous solidity.

Emilio Segré, who joined the group in 1931, describes Stern's influence on him as follows:[47]

Stern taught me a way of experimenting that I had not seen before. He calculated everything possible about his apparatus, such as the shape and intensity of the molecular beams he expected to generate, and did not proceed until preliminary experiments were in complete quantitative agreement with his calculations. This modus operandi slowed down the preliminary work, but it shortened the total time by making it possible to avoid errors and was absolutely necessary for the extremely difficult experiments Stern was conducting. The method allowed him to localize sources of misbehavior in the apparatus and of failures, and to come to a firm decision as to whether there were new and unexpected results, which occurred repeatedly. It was a rigorous and most useful schooling.

In 1959 Segré would share the Nobel Prize in Physics with Owen Chamberlain "for their discovery of the antiproton."

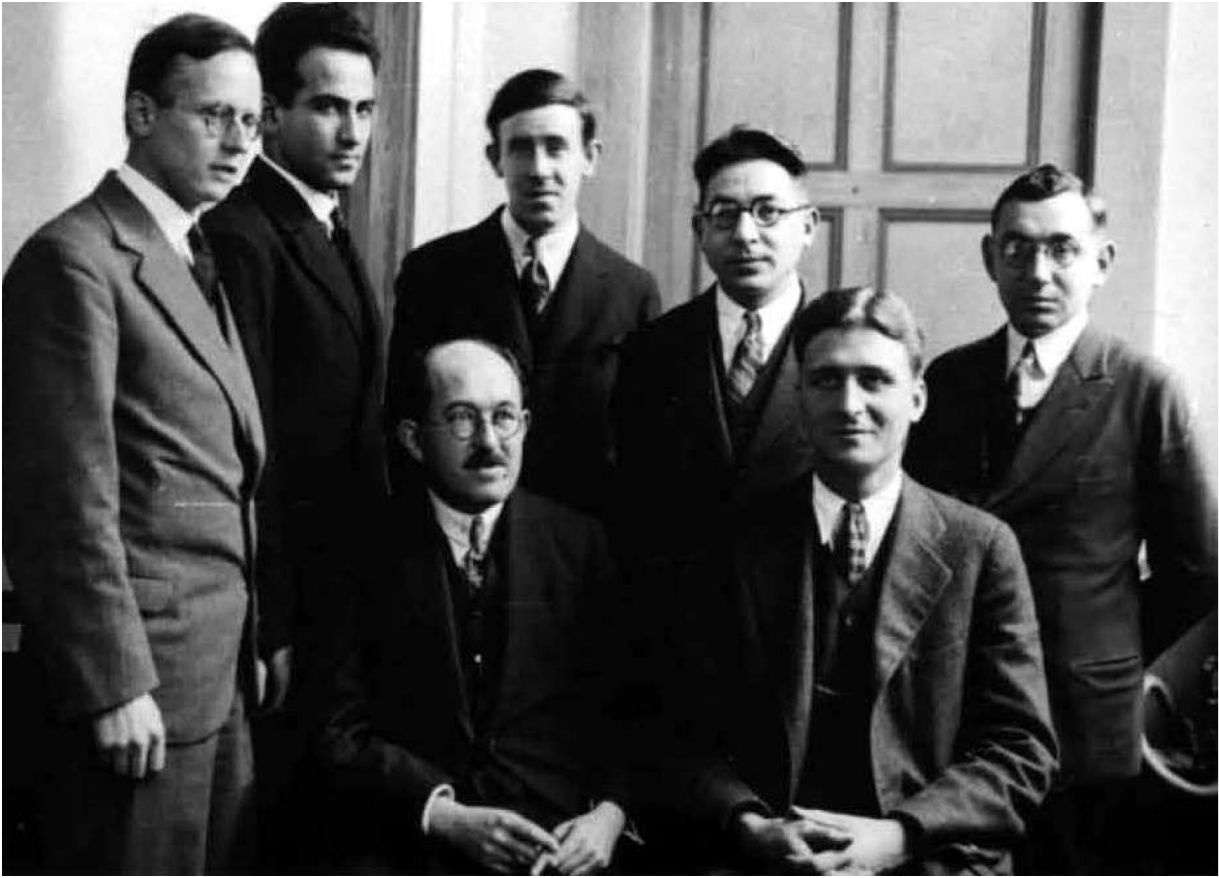

Figure 7: Stern's research group at Hamburg, 1929:[42] From left to right: Friedrich Knauer, Otto Brill, Otto Stern, Ronald Fraser, Isidor Isaac Rabi, John Bradshaw Taylor, Immanuel Estermann.

Not only were his excellent assistants and the many highly motivated fellows and students, cf. Figure 7, instrumental in fostering a stimulating and creative atmosphere in Stern's laboratory but also his outstanding university colleagues. The director of the theoretical institute was Wilhelm Lenz, Stern's predecessor in Rostock, who had assembled a group of excellent young *Privatdozents*, which included Wolfgang Pauli, Ernst Ising and Hans Jensen. Both Pauli and Jensen would receive Nobel Prizes in Physics, Pauli in 1945 "for the discovery of the Exclusion Principle, also called the Pauli Principle" and Jensen (jointly with Maria Goeppert Mayer) in 1963 "for their discoveries concerning nuclear shell structure" (this Prize was shared, in addition, with Eugene Wigner, who was cited for his work on symmetry). Wolfgang Pauli, as one of the most renowned theoreticians of the new quantum mechanics, had a strong influence on Otto Stern. Pauli who had received his PhD in 1921 under Sommerfeld, had arrived in Hamburg nearly simultaneously with Stern, following a year-long stay with Bohr in Copenhagen. In his conversation with Rigden, Rabi described the relationship between Pauli and Stern as follows:[44]

> When I was at Hamburg University, it was one of the leading centers of physics in the world. There was a close collaboration between Stern and Pauli, between experiment and theory. For example, Stern's questions were important in Pauli's theory of magnetism of free electrons in metals. Conversely, Pauli's theoretical researches were important influences in Stern's thinking. Further, Stern's and Pauli's presence attracted many illustrious visitors to Hamburg. Bohr and Ehrenfest were frequent visitors.

As Stern himself recollected in his Zürich interview, he invariably had lunch with Pauli at which time they discussed such scientific issues as "what is entropy", "how to explain the symmetry of the hydrogen molecule?" or "the problem of the zero point energy." Pauli was someone with whom he could test his "crazy" ideas.

> When I had a run of bad luck with the experiments, I would invite Pauli to dinner and pour out my troubles and invariably this helped.[4]

Since Stern was quite superstitious he was convinced that each time Pauli would enter the laboratory something would break. As a result, despite their friendship, Pauli was not permitted to enter Stern's laboratory. After Pauli left Hamburg for a professorship in Zurich in 1928 they continued their close contacts and remained lifelong friends.

Among the many pioneering experiments coming from the Hamburg laboratory, Stern considered the diffraction of atoms and molecules from surfaces to be his major contribution to the development of quantum mechanics. In his Zurich interview he explained:[4]

> I like this experiment best of all, but it is not properly understood. It has to do with the determination of the de Broglie wavelength. The apparatus consists entirely of mechanical components from the shop with the exception of the lattice constant. The atom velocity was specified with rotating slotted disks. Hitler is to be blamed that we could finish these experiments in Hamburg where it was part of our program.

In actual fact, these experiments were highly successful and were the subject of 6 publications in the UzM series. In the first of these publications with Friedrich Knauer, Stern attempted to detect the diffraction of He and $H_2$ beams from ordinary surfaces. Because of the inherent roughness of such surfaces, they realized that they would stand a chance only if they achieved a glancing angle of $10^{-3}$ radian. And indeed, in this way they succeeded in measuring reflection coefficients and even observe a hint of a diffraction peak. An attempt to diffract from an optical grating with 100 grooves per mm was, however, without success.[48] The difficulty of such an experiment can be gauged by the fact that the first such experiment would be successfully carried out only in 2010.[49] In the following spring of 1929 Stern, working alone on the very same apparatus, after some modifications finally succeeded in observing the diffraction of He and $H_2$ from a freshly cleaved NaCl crystal.[50] Then in collaboration with Immanuel Estermann and using the more inert and perfect LIF crystals, Stern obtained diffraction peaks which were sufficiently resolved to test whether they obeyed de Broglie's formula for the wavelength of matter waves. That formula had already been confirmed for electrons in 1927 by Davisson and Germer, but whether the wave-particle dualism also applied to composite particles, such as atoms or molecules, was an open issue. A year later, in a remarkable experimental tour-de-force, Estermann, Frisch and Stern succeeded in observing much sharper diffraction patterns, by first velocity-selecting the incident beams.[51] Two novel techniques were implemented for velocity selection. In one, the beam was first diffracted from a LiF crystal, which served as a grating monochromator. Only molecules scattered into a narrow angular and correspondingly narrow velocity range were then diffracted from a second LiF target crystal. In the second technique, the beam was velocity selected using two 19 cm dia. discs separated by 3.1 cm rotating on a common axis. The disks had nominally 400 equidistant radial slots each and were slightly shifted with respect to one another so that only a narrow range of velocities would be transmitted. This time the diffraction peaks were sufficiently sharp to enable quantitative testing of the de Broglie relationship, with a systematic error of only 1%, according to Stern's characteristically detailed prior analysis. Much to their disappointment, the initial value for the de Broglie wavelength deviated from the predicted one by 3%. They finally found the cause of the discrepancy as reported in a final footnote of their publication:

> The deviation was explained when after completion of the experiments the apparatus was dismantled. The slots in the velocity selector discs had been milled with a pitch circle which the manufacturer (Auerbach–Dresden) had specified to have 400 divisions on the circumference and thus we expected 400 slots. Since unfortunately we only noticed that they had 408 slots (the pitch circle was incorrectly labeled) afterwards we could then reduce the error from 3% to 1%.[51]

Thus Estermann, Frisch and Stern were the first to demonstrate the validity of the de Broglie relationship for matter waves made out of atoms and molecules.

In the course of these experiments they observed a sequence of unexpected intensity dips in the otherwise smooth diffraction peak distributions. Although unable to explain the anomalies, Frisch and Stern realized their potential significance, suspecting that they might be due to the transient adsorption of the beam particles on the surface.[52] In a later article Frisch systematically analyzed the surface components of the incident wave vector corresponding to the dips.[53] These carefully performed and documented experiments later enabled John Lennard-Jones and A.F. Devonshire to explain the dips in terms of a depletion of the scattered beam as a result of a resonant trapping into van der Waals bound states at the surface.[54] Much later in 1960's this phenomenon became known as "selective adsorption" and today is widely used to determine with high precision the attractive van der Waals potential of atoms and molecules with surfaces. Thus Stern's pioneering surface scattering experiments ultimately were continued and further perfected and would eventually become a major area of surface science research.

Two other experiments were also well before their time and pointed in directions which evolved into major fields of research only much later. Knauer in 1933 succeeded in measuring the differential cross sections for scattering of He, $H_2$, $O_2$ and $H_2O$ molecules from each other as well as $H_2$ and He from Hg atoms out to very large scattering angles.[55] Gas-phase molecular beam scattering had a renaissance in the 1960's and in 1986 lead to a Nobel Prize for Dudley Herschbach, Yuan Lee and John Polanyi "for their contributions concerning the dynamics of chemical elementary processes." In the last article of the UzM series, Frisch was able to detect the minute atomic recoil of a highly collimated sodium atom beam upon resonant photon absorption. This work was resumed in the 1970s and ultimately led to laser cooling and trapping of neutral atoms (1997 Nobel Prize in Physics, shared by Steve Chu, Claude Cohen-Tannoudji and Bill Phillips).

Most important for the later course of the then fledgling field of nuclear physics were the first measurements of nuclear magnetic moments, the most ambitious of the 8 goals proclaimed in Stern's 1926 manifesto. These were the experiments that were to garner him the Nobel Prize. Frisch and Stern in the spring of their last year in Hamburg,

with the sword of Nazism hanging over their heads[56]

finally succeeded in magnetically deflecting a beam of molecular hydrogen.[57] Even today the sensitive detection of beams of hydrogen molecules is a challenging undertaking. In their article the authors describe in meticulous detail the many modifications needed to make these experiments possible. In his 1961 Zürich interview Stern remarked:[4]

> While we were measuring the magnetic moment of the proton we were strongly chided by the theoreticians since they thought they already knew the answer.

These experiments were complicated by the fact that normal hydrogen molecules consist of 25% para-hydrogen and 75% ortho-hydrogen of which only the latter has parallel nuclear spins and a magnetic dipole moment. Since for reasons of symmetry the lowest rotational state of ortho molecules is j = 1, the interaction of the small magnetic moment associated with the rotation of the molecules (resulting from a small slippage of the electrons) with the nuclear magnetic moment also had to be accounted for. From their first deflection experiments they estimated that the ortho-component had a nuclear magnetic moment of about 2–3 nuclear magnetons ($\mu_N$) for each of the protons. A nuclear magneton is equal to the Bohr magneton, $\mu_B$, for the electron, reduced by the ratio of the electron-to-proton mass, i.e., $\mu_N = \mu_B/1836$. According to the then prevalent theory due to Dirac of particles with spin ½,[58] the magnetic moment of the proton should have been equal to $\mu_N$. Hence Stern's result was in clear contradiction with theory and implied that the proton had an inner structure. Less than two months later, Estermann and Stern repeated the measurements and reported a value of 2.5 $\mu_N$ for protons with an error of only 10%,[59] which is within their error bars consistent with the present-day value of 2.7896 $\mu_N$. They also reported a value of 0.8–0.9 nuclear magnetons for the rotational magnetic moment, obtained from the deflection of a specially prepared beam of pure para-hydrogen, in excellent agreement with the presently accepted value of 0.88291 $\mu_N$.

The numerous spectacular achievements of Stern's group were documented in 45 publications, including the 30 in the UzM series. Hamburg was an international hub of physics, which had attracted many distinguished visitors. Stern had been invited to numerous national and international meetings. In connection with the importance of his year in Hamburg Rabi told Rigden:[44]

> From Stern and from Pauli I learned what physics should be. For me it was not a matter of more knowledge. … Rather it was the development of taste and insight; it was the development of standards to guide research, a feeling for what is good and what is not good.

Stern had this quality of taste in physics and he had it to the highest degree. As far as I know, Stern never devoted himself to a minor problem.

Shortly after leaving Hamburg in early 1929, Rabi was offered a lectureship at Columbia University. Although intending to do theory he soon revived his interest in magnetic moments of the nuclei. After trying a number of different magnetic field arrangements with his students S. Millman, Polykarp Kusch and Jerrold Zacharias, Rabi finally in 1937 devised a new scheme for measuring nuclear magnetic moments, illustrated in Fig. 8.

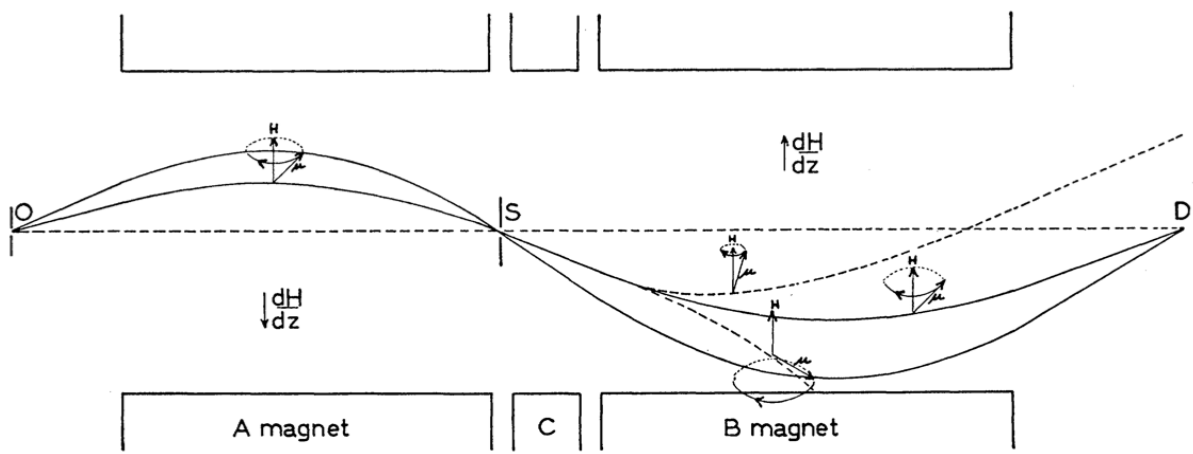

Figure 8: Schematic diagram of Rabi's 1939 apparatus for measuring nuclear magnetic moments.[60]

Instead of using only one magnetic field as in the Stern-Gerlach apparatus, now two identical inhomogeneous magnets denoted the A and B magnets were introduced. In sector C a radio frequency electromagnetic field was applied to induce transitions from one hyperfine state to another and a superimposed homogeneous magnet insured that the spin orientation was not lost in passing from one magnet to the next. The advantage of this arrangement is, as illustrated by the solid line trajectories in Fig. 8, that atoms with a wide range of velocities, which are focused by magnet A on the slit *s*, are refocused on the detector at D by magnet B. As a result of a transition induced by the radio frequency field C to an to an atomic state with a different magnetic moment, the affected atoms undergo a different deflection in the B magnet (dashed lines in Fig. 7); since such atoms no longer arrive at the detector, a decrease in the detector signal is observed. Rabi called this new technique "Molecular Beam Resonance Method for Measuring Nuclear Magnetic Moments." This method was soon adopted by many other groups world-wide. It has the advantage of combining the high precision of spectroscopy with the state selectivity of the molecular beam method, independent of the beam's velocity spread. This development brought Rabi the 1944 Nobel

Prize in Physics and later to several of his former students, including Polykarp Kusch (1955), Norman Ramsey (1989) and Charles Townes (1964). It also led to the development of the nuclear magnetic resonance spectroscopy of solids and biomolecules and nuclear magnetic resonance imaging in medicine.

In 1939 Rabi and his coworkers succeeded in a precision measurement of the nuclear magnetic dipole moment of the proton: $2.78 \pm 0.02 \ \mu_N$ and of the deuteron: $0.853 \pm 0.007 \ \mu_N$ and were able to establish that nuclei have a quadrupole moment. Thus not only did they confirm and perfect the early measurements of Frisch and Stern, but they also introduced a new powerful method for the spectroscopy of nuclei.

On April 7, 1933, the Nazis promulgated the *Gesetz zur Wiederherstellung des Berufsbeamtentums* (Law for the Restoration of Professional Civil Service), which served to enforce the political conformity of civil servants, including university employees, and formed an early peak in the persecution and disenfranchisement of citizens of Jewish descent in Germany.[61] The law provided a basis for the ouster of all of Stern's Jewish coworkers: Estermann, Schurmann and also Frisch – even though he was an Austrian citizen – received their letters of dismissal on June 23, 1933. Since Stern had served in the First World War, he was exempted from the law. However, it had become clear that more anti-semitic legislation and other discriminatory measures were to follow. Therefore, Stern submitted his letter of resignation just a few days later, on June 30, 1933. His resignation was to take effect on October 1, 1933. Before leaving Germany, Stern made sure that Friedrich Knauer could complete his habilitation. The final paper of the UzM series, on photon recoil, authored by Otto Robert Frisch and submitted on August 22, 1933, closed with the following statement:[62]

> It would have been possible to obtain significantly improved results, but the experiments had to be prematurely terminated for external reasons.

The expulsion of Otto Stern and his coworkers from their posts at Hamburg and ultimately from Germany is among the most manifest examples of the injustice and insanity of Nazi pre-war policies toward the Jewish members of German academia.[63]

### Emigration to the U.S.A. in 1933

Stern was more fortunate than many of his Jewish contemporaries in that he was offered a research professorship together with his longtime colleague Immanuel Estermann at the Carnegie Institute in Pittsburgh. However, they were disappointed by the poor conditions they found in Pittsburgh. Estermann wrote:

> The support provided for Stern during the depression was quite meager. Stern was unable to regain the drive of the Hamburg laboratory although a number of important publications originated at the Carnegie Institute.

Six publications appeared in the 12 years until 1945 that Stern had spent in Pittsburgh but the significance of any of them would not come even close to the significance of the papers produced in Frankfurt and Hamburg. In the U.S. Stern was in great demand as a speaker. Already in 1930 he received an honorary degree from Berkeley and in 1936 he was invited to become a member of the Royal Danish Academy. On March 8, 1939 Stern became a U.S. citizen enabling him to participate in secret military research projects. On November 9, 1944 he was informed that he would receive an unshared Nobel Prize in Physics (for the year 1943), a recognition which was indeed long overdue.

As noted in the Introduction, Otto Stern was, with eighty one nominations for a Nobel Prize in Physics, the most nominated candidate between 1901 and 1950. Only Arnold Sommerfeld, who did not receive the Prize, had nearly as many nominations (80). Stern's nominators were James Franck, Max Planck, Albert Einstein, Niels Bohr, Max Born, Willy Wien, Johannes Stark, Pierre Weiss, Max von Laue, Chandrasekhara Venkata Raman, Oscar Klein, Werner Heisenberg, Friedrich Hund, Wolfgang Pauli, Gregor Wentzel, Peter Pringsheim, Rudolf Ladenburg, Eugen Wigner, Carl David Anderson, Manne Siegbahn, Arthur Compton, Hans Bethe, and many others. In 1934 and 1940 the Nobel Prize in Physics was not awarded even though Stern had been nominated 15 and 14 times, respectively. The reason put forth by the five-member Nobel Committee behind the slow coming of the Prize for Stern was that space quantization was nothing fundamentally new since it had been predicted by Sommerfeld already in 1916. Moreover, Stern's value for the magnetic moment of the proton of 2.5 disagreed with other published results. Dirac's rudimentary theory had predicted a value of 1, Landé a value of 2,[64] and Rabi had in 1934 reported a value of 3.25 $\mu_N$ with a 10% error. But Stern's 1933 value had in fact come closest to the present-day accepted value. Stern's former Hamburg postdoctoral fellow and friend Isidor Rabi received the Nobel Prize for Physics for the subsequent year 1944. They must have had a happy reunion on December 10, 1944 in New York City at the Waldorf Astoria Hotel where, because of the ongoing war, the Nobel ceremony had taken place. The Nobel medals were bestowed on both of them and other laureates by the Swedish ambassador Eric Boström. In his Nobel Lecture, delivered in 1946, Stern extolled the molecular beam method:[65]

> The most distinctive characteristic property of the molecular ray method is its simplicity and directness. It enables us to make measurements on isolated neutral atoms or molecules with macroscopic tools. For this reason it is especially valuable for testing and demonstrating directly fundamental assumptions of the theory.

Interestingly, Stern himself had submitted only two Nobel nominations (up until 1950):**Error! Bookmark not defined.** in 1933 of Gilbert N. Lewis (unsuccessful) and in 1949 of Hidei Yukawa (successful).

In 1945-1946 Stern retired from the Carnegie Institute and moved to Berkeley where some of his close relatives had lived. The bachelor Otto Stern bought a house on Cragmont Avenue with a nice view of the San Francisco Bay. There he planned to live with his unmarried younger sister Elise, but she died unexpectedly in 1945. His elder sister Berta lived there as well, with her husband Walter Joseph Kamm and their children. As Emilio Segré reports in his brief biographical article, Stern was a frequent visitor in the colloquia and seminars given at the University of California at Berkeley.

After the war Stern generously helped many of his friends with CARE packages. He supported von Laue even with clothing, since von Laue had lost his property when his house was bombed during the war. Their rich correspondence revolved about everyday issues as well as bigger themes, such as Stern's relation to his former homeland. In a letter from October 1, 1947, von Laue pointed out that:

> We all must throw our resentments [about the wrongs suffered during the Nazi era – however understandable] overboard, if human kind should be saved from going under.

As implied by his correspondence with Lise Meitner, Max von Laue and Hans Jensen, Stern would not miss an opportunity to visit Europe – to see his friends at conferences and meetings, in particular in Copenhagen, London, and foremost in Zurich. Stern would visit Zurich almost every year for a period of several months and usually stay in pension *Tiefenau* at *8 Steinwiesenstrasse*. Nearly always Stern would cross the ocean by ship. When he arrived by ship in England or Holland he would on occasion pass through Germany en route to Zurich. Despite being deeply rooted in German culture, Stern took pains to meet his German friends outside of Germany. He invited several of them at his own expense on holidays in Zurich, but he never again "officially" visited Germany. In the 1950s, on his only private trip to Germany, he visited his friend Max Volmer in East Berlin in the GDR. After the war he turned down the offer by the city of Hamburg to pay him his life annuity and never accepted the offers by von Laue and others to become a member of the Academy of Sciences in Göttingen. In the summer of 1968, Stern attended the annual Nobel Laureate meeting in Lindau. On August 17, 1969, he suffered a heart attack while in a cinema in Berkeley and died a few days later in a hospital. According to Peter Toschek's article,[66] Stern once remarked during his Hamburg golden years that

> it would be nice to die while watching a good movie. […] Whether the movie was good or not has not been passed on.

Stern's ashes have been buried at the "Sunset View Cemetery" in El Ceritto near Berkeley.

Otto Stern was known for his kind and gentlemanly personality. Rabi wrote in his obituary for Stern:[67]

> Stern was one of the antistuffy generation of German professors who observed with a mixture of amusement and contempt the pomposity of their predecessors.

Or, as a U.S. newspaper put it:[68]

> Looks – the jovial type, fine smile, personality and temperament. He has the best traits of the European gentleman.

Gerlach survived Otto Stern by about 10 years. When Stern died, Gerlach wrote in the *Physikalische Blätter*:[18]

> Those of us who knew Stern valued his openness – he was a Grand seigneur! – His absolute reliability, his often spontaneous but not always conciliatory (*einfach*) but fruitful discussions and for those, who had a sense for such, his even sarcastic but always well considered judgment about objects and persons. He deplored arrogance and bad manners. Although trained as a theoretician he was full of ideas for experiments, never at a loss for a new suggestion, when the first attempt failed.

The German Physical Society honored Otto Stern's and Walther Gerlach's legacy in 1992 by establishing in parallel to the existing Max Planck Medal for excellence in theory a new distinction, "The Stern-Gerlach Medal," for excellence in experimental physics.

**Acknowledgements:**

We are grateful to the Stern family (Liselotte and David Templeton as well as Diana Templeton-Killen and Alan Templeton) for their private communications as well as for providing invaluable documents and photographs to us. We also acknowledge the help we received from the archives in Berkeley (Bancroft library, David Farrel), in Frankfurt (Goethe Universität Frankfurt, Wolfgang Trageser and Michael Maaser) and in Stockholm (Nobelarchive, Carl Grandin). Additional help was generously provided by the Institut für Physikalische Chemie of the Universität Hamburg (Horst Förster und Fritz Thieme) as well as by Karin Reich (Universität Hamburg) and Bruno Lüthi (Goethe Universität Frankfurt and ETH Zürich). We also thank our Hamburg colleagues Volkmar Vill and Peter Toschek for helpful correspondence.